# Electron-phonon coupling from *GW* perturbation theory: Practical workflow combining BerkeleyGW, ABINIT, and EPW

Zhenglu Li[1,2,3], Gabriel Antonius[1,2,5], Yang-Hao Chan[1,2,4], and Steven G. Louie[1,2,*]

[1]Department of Physics, University of California at Berkeley, Berkeley, California 94720, USA

[2]Materials Sciences Division, Lawrence Berkeley National Laboratory, Berkeley, California 74720, USA

[3]Mork Family Department of Chemical Engineering and Materials Science, University of Southern California, Los Angeles, California 90089, USA

[4]Institute of Atomic and Molecular Sciences, Academia Sinica and Physics Division, National Center of Theoretical Sciences, Taipei 10617, Taiwan

[5]Present Address: Département de Chimie, Biochimie et Physique, Institut de recherche sur l'hydrogène, Université du Québec à Trois-Rivières, C.P. 500, Trois-Rivières, Canada

[*]E-mail: sglouie@berkeley.edu

We present a workflow of practical calculations of electron-phonon (*e*-ph) coupling with many-electron correlation effects included using the *GW* perturbation theory (*GW*PT). This workflow combines BerkeleyGW, ABINIT, and EPW software packages to enable accurate *e*-ph calculations at the *GW* self-energy level, going beyond standard calculations based on density functional theory (DFT) and density-functional perturbation theory (DFPT). This workflow begins with DFT and DFPT calculations (ABINIT) as starting point, followed by *GW* and *GW*PT calculations (BerkeleyGW) for the quasiparticle band structures and *e*-ph matrix elements on coarse electron **k**- and phonon **q**-grids, which are then interpolated to finer grids through Wannier interpolation (EPW) for computations of various *e*-ph coupling determined physical quantities such as the electron self-energies or solutions of anisotropic Eliashberg equations, among others. A gauge-recovering symmetry unfolding technique is developed to reduce the computational cost of *GW*PT (as well as DFPT) while fulfilling the gauge consistency requirement for Wannier interpolation.

## I. Introduction

First-principles calculations of electron-phonon (*e*-ph) coupling have become a standard, essential approach in understanding and predicting a wide range of phenomena in real materials including electronic and thermal transport, phonon-assisted optical absorption, phonon-mediated superconductivity, electron and phonon spectral functions, to name a few [1]. The core building blocks of *e*-ph theories for different phenomena are the electron band energies and *e*-ph matrix elements [1]. To date, a prevailing and successful *ab initio* approach for accurate *e*-ph coupling calculations is through the combination of density functional theory (DFT) and density-functional perturbation theory (DFPT), along with Wannier interpolation [1-11]. However, growing evidence shows that the exchange-correlation potential in DFT (used to determine the Kohn-Sham orbital energies) sometimes fails to properly capture the self-energy effects on the real quasiparticles, leading to inaccuracy in describing their energies [12-14] and *e*-ph matrix elements [15-23].

The recent development of *GW* perturbation theory (*GW*PT) [21], along with the well-established *GW* method [12,13,24,25], allows for first-principles *e*-ph computation at the many-electron level, where the self-energy effects are consistently formulated within the *GW* approximation and included in both the electron band energies and the *e*-ph matrix elements [21-23]. Using the *GW* and *GW*PT approaches beyond the widely-used DFT and DFPT methods, we have revealed strong correlation-enhancement effects in the *e*-ph coupling of several oxide superconductors, namely, $Ba_{1-x}K_xBiO_3$ [21], $La_{2-x}Sr_xCuO_4$ [22], and $Nd_{1-x}Sr_xNiO_2$ [23]. The computation of many *e*-ph phenomena however requires fine sampling of the electron and phonon states in the Brillouin zone (BZ), to which the Wannier interpolation of materials' electronic structure and *e*-ph matrix elements provides an effective solution [4]. We have developed here a workflow that connects BerkeleyGW [26,27], ABINIT [28,29], and EPW [7], enabling the Wannier interpolation (using EPW) of *GW* electron states and *GW*PT *e*-ph matrix elements (from BerkeleyGW), as well as those of DFT and DFPT (from ABINIT). Consequently, accurate *ab initio* *e*-ph calculations at the many-electron *GW* level can now be carried out with significant increase in computational ease.

In this paper, we present details of this workflow for practical *e*-ph calculations at the *GW* level, in particular with *GW*PT. Like standard *GW* calculations which use DFT outputs as the starting point, *GW*PT calculations use DFPT outputs as the starting point. A wrapper code takes DFT-DFPT outputs from ABINIT [29] and *GW*-*GW*PT outputs from BerkeleyGW [26], and then prepares data into a specific format for a modified version of EPW [7] to read. The *e*-ph properties then are computed using the interpolated electron states and *e*-ph matrix elements with EPW. To reduce the computational cost using the symmetry of the crystal, the *e*-ph matrix elements from DFPT and *GW*PT need only be computed on the symmetry-reduced **q**-grid (and full **k**-grid). A gauge-recovering symmetry unfolding technique is developed to fulfill the gauge consistency requirement for the Wannier interpolation procedure. This workflow not only enables the first-principles *e*-ph calculations at the *GW*-*GW*PT level with Wannier interpolation, but also extends the interface options of EPW going beyond the use of Quantum ESPRESSO [30,31] by including ABINIT and BerkeleyGW, and in general any software packages that produce *e*-ph matrix elements.

## II. *GW*PT calculations with DFPT as start point

The recently developed *GW*PT method within the linear-response framework enables the systematic computation of *e*-ph matrix elements at the *GW* level [21]. The theoretical essence and some applications of *GW*PT have been discussed in Refs. [21-23], and the detailed derivations of the theory and extended discussions of different aspects of *GW*PT will be published elsewhere [32]. Here, we focus on the practical workflow of *GW*PT calculations. We shall also restrict, in this paper, to the formalism and calculations of non-magnetic systems, i.e., time-reversal symmetry is present, and no spin polarizations nor spin-orbit coupling are included.

The linear-response framework of DFPT and $GW$PT decouples the different phonon modes (labeled by the phonon wavevector **q** and branch index $\nu$), and the responses (changes in physical quantities) to each mode can be computed within a single primitive unit cell. $GW$PT computes all needed $e$-ph matrix elements at the $GW$ level, denoted in the phonon-mode basis as [21],

$$g_{mn\nu}^{GW}(\mathbf{k},\mathbf{q}) = \langle\psi_{m\mathbf{k+q}}|\Delta_{\mathbf{q}\nu}V_{\text{ion}}|\psi_{n\mathbf{k}}\rangle + \langle\psi_{m\mathbf{k+q}}|\Delta_{\mathbf{q}\nu}V_{\text{H}}|\psi_{n\mathbf{k}}\rangle + \langle\psi_{m\mathbf{k+q}}|\Delta_{\mathbf{q}\nu}\Sigma|\psi_{n\mathbf{k}}\rangle, \quad (1)$$

where $m$ and $n$ label the electron bands, **k** and **q** are wavevectors, $V_{\text{ion}}$ is the ionic potential (pseudopotential), $V_{\text{H}}$ is the Hartree potential, and $\Sigma = iGW$ is the electron self-energy in the $GW$ approximation. Through DFPT, the DFT $e$-ph matrix elements is computed with [1],

$$\begin{aligned}g_{mn\nu}^{\text{DFT}}(\mathbf{k},\mathbf{q}) &= \langle\psi_{m\mathbf{k+q}}|\Delta_{\mathbf{q}\nu}V^{\text{KS}}|\psi_{n\mathbf{k}}\rangle \\ &= \langle\psi_{m\mathbf{k+q}}|\Delta_{\mathbf{q}\nu}V_{\text{ion}}|\psi_{n\mathbf{k}}\rangle + \langle\psi_{m\mathbf{k+q}}|\Delta_{\mathbf{q}\nu}V_{\text{H}}|\psi_{n\mathbf{k}}\rangle + \langle\psi_{m\mathbf{k+q}}|\Delta_{\mathbf{q}\nu}V_{\text{xc}}|\psi_{n\mathbf{k}}\rangle,\end{aligned} \quad (2)$$

where $V^{\text{KS}}$ is the Kohn-Sham potential, and $V_{\text{xc}}$ is the exchange-correlation potential. Comparing the two equations, $g_{mn\nu}^{GW}(\mathbf{k},\mathbf{q})$ can be practically constructed by replacing the $V_{\text{xc}}$ contribution by the self-energy contribution [21],

$$g_{mn\nu}^{GW}(\mathbf{k},\mathbf{q}) = g_{mn\nu}^{\text{DFT}}(\mathbf{k},\mathbf{q}) - \langle\psi_{m\mathbf{k+q}}|\Delta_{\mathbf{q}\nu}V_{\text{xc}}|\psi_{n\mathbf{k}}\rangle + \langle\psi_{m\mathbf{k+q}}|\Delta_{\mathbf{q}\nu}\Sigma|\psi_{n\mathbf{k}}\rangle. \quad (3)$$

In the above equations, the differential operator $\Delta_{\mathbf{q}\nu}$ in the phonon mode basis is defined as [1,21],

$$\Delta_{\mathbf{q}\nu} = \sqrt{\frac{\hbar}{2\omega_{\mathbf{q}\nu}}} \sum_{\kappa\alpha} \frac{1}{\sqrt{M_\kappa}} e_{\kappa\alpha,\nu}(\mathbf{q}) \sum_l^{N_l} e^{i\mathbf{q}\cdot\mathbf{R}_l} \frac{\partial}{\partial\tau_{\kappa\alpha l}}, \quad (4)$$

where $\omega_{\mathbf{q}\nu}$ is the eigenfrequency of the phonon mode $\mathbf{q}\nu$, $\kappa$ labels the atoms within a unit cell, and $l$ labels unit cells (with Born-von Karman boundary conditions), $M_\kappa$ is the atomic mass, $\alpha = x, y, z$ labels the Cartesian directions, $\tau_{\kappa\alpha l}$ is an basis atom coordinate, and $e_{\kappa\alpha,\nu}(\mathbf{q})$ is the $\kappa\alpha$ component of the phonon eigenvector $e_{\mathbf{q}\nu}$. In practice, without prior knowledge of phonon frequencies and eigenvectors, DFPT typically solves for the responses to static atom perturbations along Cartesian directions or lattice vectors, rather than in the phonon eigen-mode basis. ABINIT uses differential operators with respect to periodic atom displacements along the three primitive unit cell lattice vectors, which are defined as,

$$\Delta_{\mathbf{q}\kappa a} = \sum_l^{N_l} e^{i\mathbf{q}\cdot\mathbf{R}_l} \frac{\partial}{\partial\tau_{\kappa a l}}, \quad (5)$$

where $a = \mathbf{a}_1, \mathbf{a}_2, \mathbf{a}_3$ labels the lattice vectors, and we denote this reference frame as in crystal coordinates. The $GW$PT implementation follows the same convention as in ABINIT.

Standard first-principles $GW$ codes, such as BerkeleyGW, use eigenvalues and wavefunctions of DFT as a starting point to construct and compute the quasiparticle self-energy operator $\Sigma$. Similarly, for $GW$PT, outputs of DFPT - including $g_{mn\kappa a}^{\text{DFT}}(\mathbf{k},\mathbf{q})$ and the first-order change in the wavefunctions $\Delta_{\mathbf{q}\kappa a}\psi_{n\mathbf{k}}$ - are used for computing $g_{mn\kappa a}^{GW}(\mathbf{k},\mathbf{q})$. The complete first-order change in the wavefunction at the DFT level reads,

$$\Delta_{\mathbf{q}\kappa a}\psi_{n\mathbf{k}}(\mathbf{r}) = \sum_m \frac{\langle\psi_{m\mathbf{k+q}}|\Delta_{\mathbf{q}\kappa a}V^{\text{KS}}|\psi_{n\mathbf{k}}\rangle}{\varepsilon_{n\mathbf{k}} - \varepsilon_{m\mathbf{k+q}}} \psi_{m\mathbf{k+q}}(\mathbf{r}) = \sum_m \frac{g_{mn\kappa a}^{\text{DFT}}(\mathbf{k},\mathbf{q})}{\varepsilon_{n\mathbf{k}} - \varepsilon_{m\mathbf{k+q}}} \psi_{m\mathbf{k+q}}(\mathbf{r}), \quad (6)$$

where $\varepsilon_{n\mathbf{k}}$ is the Kohn-Sham DFT eigenvalues, and the summation over the band index $m$ includes the full Hilbert space (defined by the wavefunction energy cutoff for a planewave basis set). Standard DFPT implementations solve a Sternheimer equation to obtain the first-order change in the wavefunctions, but project $\Delta_{\mathbf{q}\kappa a}\psi_{n\mathbf{k}}(\mathbf{r})$ on the subspace of certain unoccupied states [33]. However, GWPT calculations require the complete form of $\Delta_{\mathbf{q}\kappa a}\psi_{n\mathbf{k}}(\mathbf{r})$ defined in Eq. (6), without any restriction on the band index $m$. We have thus enabled ABINIT to construct and output the complete $\Delta_{\mathbf{q}\kappa a}\psi_{n\mathbf{k}}(\mathbf{r})$ on demand. In the above equation, we introduced the e-ph matrix elements in the crystal-coordinate basis,

$$g_{mn\kappa a}^{\text{DFT}}(\mathbf{k},\mathbf{q}) = \langle\psi_{m\mathbf{k}+\mathbf{q}}|\Delta_{\mathbf{q}\kappa a}V^{\text{KS}}|\psi_{n\mathbf{k}}\rangle, \quad (7)$$

which can be directly computed in ABINIT. Moreover, the first-order change in the exchange-correlation potential $\Delta_{\mathbf{q}\kappa a}V_{\text{xc}}(\mathbf{r})$ can be readily obtained from DFPT, and therefore its corresponding matrix elements are computed as,

$$g_{mn\kappa a}^{\text{xc}}(\mathbf{k},\mathbf{q}) = \langle\psi_{m\mathbf{k}+\mathbf{q}}|\Delta_{\mathbf{q}\kappa a}V_{\text{xc}}|\psi_{n\mathbf{k}}\rangle. \quad (8)$$

Eqs. (6) – (8) are quantities imported from DFPT for the GWPT calculations.

The main workload of GWPT is to construct the first-order change in the GW self-energy operator and to evaluate its matrix elements. With the wavefunctions $\psi_{n\mathbf{k}}(\mathbf{r})$ from DFT and their first-order changes $\Delta_{\mathbf{q}\kappa a}\psi_{n\mathbf{k}}(\mathbf{r})$ from DFPT, the first-order change in the Green's function can be constructed as [21],

$$\Delta_{\mathbf{q}\kappa a}G(\mathbf{r},\mathbf{r}';\varepsilon) = \sum_{n\mathbf{k}}\frac{\Delta_{\mathbf{q}\kappa a}\psi_{n\mathbf{k}}(\mathbf{r})\psi_{n\mathbf{k}}^{*}(\mathbf{r}') + \psi_{n\mathbf{k}}(\mathbf{r})[\Delta_{-\mathbf{q}\kappa a}\psi_{n\mathbf{k}}(\mathbf{r}')]^{*}}{\varepsilon - \varepsilon_{n\mathbf{k}} - i\delta_{n\mathbf{k}}}, \quad (9)$$

where $\delta_{n\mathbf{k}} = 0^+$ (or $0^-$) for $\varepsilon_{n\mathbf{k}} < \varepsilon_F$ (or $\varepsilon_{n\mathbf{k}} > \varepsilon_F$) and $\varepsilon_F$ is the Fermi energy. Here, the summation over the band index $n$ includes both occupied and unoccupied states, and a large number of bands is typically needed for the convergence of the self-energy and its first-order changes. We further take a constant-screening approximation by neglecting the responses in the screened Coulomb interaction, i.e., $\Delta_{\mathbf{q}\kappa a}W = 0$ [19,21], which is expected to be a good approximation in semiconductors (insulators) and metals where the screening environment is robust against phonon perturbations. This approximation is equivalent to the standard and well-justified approximation of $\frac{\delta W}{\delta G} = 0$ in constructing the electron-hole kernel in the GW-Bethe-Salpeter equation (GW-BSE) approach [34,35]. Consequently, the first-order change in the GW self-energy operator is written, in the frequency domain, as [21],

$$\Delta_{\mathbf{q}\kappa a}\Sigma(\mathbf{r},\mathbf{r}';\varepsilon) = i\int\frac{d\varepsilon'}{2\pi}e^{-i\delta\varepsilon'}\Delta_{\mathbf{q}\kappa a}G(\mathbf{r},\mathbf{r}';\varepsilon-\varepsilon')W(\mathbf{r},\mathbf{r}';\varepsilon'), \quad (10)$$

where $\delta = 0^+$. BerkeleyGW directly computes the corresponding matrix elements,

$$g_{mn\kappa a}^{\Sigma}(\mathbf{k},\mathbf{q}) = \langle\psi_{m\mathbf{k}+\mathbf{q}}|\Delta_{\mathbf{q}\kappa a}\Sigma|\psi_{n\mathbf{k}}\rangle. \quad (11)$$

The e-ph matrix elements at the GW level are constructed in practice as,

$$g_{mn\kappa a}^{GW}(\mathbf{k},\mathbf{q}) = g_{mn\kappa a}^{\text{DFT}}(\mathbf{k},\mathbf{q}) - g_{mn\kappa a}^{\text{xc}}(\mathbf{k},\mathbf{q}) + g_{mn\kappa a}^{\Sigma}(\mathbf{k},\mathbf{q}). \quad (12)$$

Using the phonon frequencies and eigenvectors, e-ph matrix elements in the crystal-coordinate basis can be rotated to the phonon-mode basis as in Eq. (1).

### III. ABINIT-BerkeleyGW-EPW interface wrapper

In this section, we present a workflow enabled by a wrapper code elph_interface for connecting ABINIT and BerkeleyGW with the EPW code, which is a popular software package for performing Wannier interpolation of e-ph matrix elements. The standard public version of the EPW code thus far is distributed and interfaced with the Quantum Espresso package. Our workflow allows us not only to interpolate $GW$PT $e$-ph matrix elements $g^{GW}_{mn\nu}(\mathbf{k},\mathbf{q})$ from BerkeleyGW using the EPW code, but also to connect EPW with ABINIT for the interpolation of DFPT $e$-ph matrix elements.

Fig. 1 shows the whole workflow. The workflow starts with the DFT calculations using ABINIT, producing DFT electronic structure of the material system being studied. One central quantity from the DFT step is a single set of wavefunctions $\{\psi_{n\mathbf{k}}\}$ uniformly sampling the full $\mathbf{k}$-BZ. Because the $e$-ph matrix elements $g^{\text{DFT}}_{mn\kappa a}(\mathbf{k},\mathbf{q})$ and $g^{GW}_{mn\kappa a}(\mathbf{k},\mathbf{q})$, the first-order change in the wavefunctions $\Delta_{\mathbf{q}\kappa a}\psi_{n\mathbf{k}}$, and the Wannier transformations are all gauge-dependent, these quantities are required to be fixed to the particular gauge unique to this set of wavefunctions $\{\psi_{n\mathbf{k}}\}$. In Fig. 1, gauge-consistent quantities (defined to be quantities need to have the same gauge in the computation) are highlighted in the green boxes. Note that this set of wavefunctions $\{\psi_{n\mathbf{k}}\}$ should have enough empty bands for converged $GW$ and $GW$PT calculations. With $\{\psi_{n\mathbf{k}}\}$ on the full $\mathbf{k}$-grid, Wannierization [36-38] can be performed using the Wannier90 package [39,40] to generate the Wannier transformation matrix $U_{nw\mathbf{k}}$, where $w$ labels the Wannier basis functions. The $U_{nw\mathbf{k}}$ matrix will be used later at the EPW step. With the implementation of DFPT in ABINIT, phonon frequencies $\omega_{\mathbf{q}\nu}$ and eigenvectors $e_{\mathbf{q}\nu}$ (or equivalently, dynamical matrices), $e$-ph matrix elements $g^{\text{DFT}}_{mn\kappa a}(\mathbf{k},\mathbf{q})$, first-order changes in wavefunctions $\Delta_{\mathbf{q}\kappa a}\psi_{n\mathbf{k}}$, and exchange-correlation potentials $\Delta_{\mathbf{q}\kappa a}V_{\text{xc}}$ can be computed (in the crystal-coordinate basis).

The $GW$ and $GW$PT calculations both need the inverse dielectric matrix $\epsilon^{-1}_{\mathbf{GG}'}(\mathbf{p})$, where $\mathbf{G}$ and $\mathbf{G}'$ are the reciprocal lattice vectors, and $\mathbf{p}$ is a wavevector for internal summation in constructing the self-energy operator and its derivatives. In BerkeleyGW, $\epsilon^{-1}_{\mathbf{GG}'}(\mathbf{p})$ is constructed using the executable epsilon.cplx.x (complex flavor required in BerkeleyGW for $GW$PT) within the random-phase approximation. Its frequency dependence is treated with the Hybertsen-Louie plasmon-pole model [13] for the current $GW$PT implementation. The executable sigma.cplx.x performs $GW$ and $GW$PT calculations, computing the $GW$-level quasiparticle energies $\varepsilon^{GW}_{n\mathbf{k}}$ and $e$-ph matrix elements $g^{GW}_{mn\kappa a}(\mathbf{k},\mathbf{q})$, respectively. The wrapper code elph_interface then postprocesses the data from DFPT and $GW$PT. Symmetry reduction of the $\mathbf{q}$-mesh can be utilized for $GW$PT (as well as DFPT). The full $e$-ph matrix elements on the full $\mathbf{q}$-BZ can be obtained by an unfolding process using the executable sympert.x within BerkeleyGW, as discussed in the next section. Finally, elph_interface prepares input data - following the EPW convention (formats and units) – to be read by an in-house modified version of EPW (v4), and in particular, the $e$-ph matrix elements are rotated into Cartesian-coordinate basis $g^{\text{DFT}}_{mn\kappa\alpha}(\mathbf{k},\mathbf{q})$ and $g^{GW}_{mn\kappa\alpha}(\mathbf{k},\mathbf{q})$.

Within our workflow, the Wannierization using Wannier90 code is performed as a separate step. The modified EPW skips its original reading of DFT wavefunctions and calculations of $g^{\text{DFT}}_{mn\kappa\alpha}(\mathbf{k},\mathbf{q})$, but instead it directly reads in all precomputed data (from $GW$PT with BerkeleyGW or DFPT with ABINIT) prepared by elph_interface. All symmetries are disabled in the modified EPW. Using Wannier functions, EPW interpolates the electronic structure and $e$-ph matrix elements from the coarse $\mathbf{k}$- and $\mathbf{q}$-meshes to fine $\mathbf{k}$- and $\mathbf{q}$-meshes or to arbitrary $\mathbf{k}$- and $\mathbf{q}$-points, as well as rotating the $e$-ph matrix elements into the phonon-mode basis. Then subsequent EPW calculations can be performed as usual. The interpolated $g^{\text{DFT}}_{mn\nu}(\mathbf{k}_{\text{fi}},\mathbf{q}_{\text{fi}})$ and $g^{GW}_{mn\nu}(\mathbf{k}_{\text{fi}},\mathbf{q}_{\text{fi}})$ can be used for computing many $e$-ph properties, such as $e$-ph induced electron self-energy $\Sigma^{e-\text{ph}}_{n\mathbf{k}}(\omega)$, electron spectral function $A(\mathbf{k},\omega)$, phonon self-energy $\Pi^{e-\text{ph}}_{\mathbf{q}\nu}(\omega)$, electric and thermal transport coefficients, and phonon-mediated superconductivity via the anisotropic Eliashberg equations, among others [1].

**IV. Gauge-recovering symmetry unfolding technique for $e$-ph matrix elements**

Due to the heavy computational expense of $GW$PT in computing $g_{mn\nu}^{GW}(\mathbf{k}, \mathbf{q})$, symmetry reduction in the number of matrix elements needed from direct calculations is paramount. However, a complication arises when performing a Wannier interpolation of *e*-ph matrix elements obtained on a symmetry-reduced **q**-mesh. The Wannierization procedure with wavefunctions on a full **k**-grid generates a basis transformation matrix with its gauge fixed to this specific set of wavefunctions $\{\psi_{n\mathbf{k}}\}$. As illustrated in Fig. 2 with a symmetry-reduced **q**-mesh [Fig. 2(a)], the unfolding process (for wavefunctions and *e*-ph matrix elements) generates a different gauge at the rotated **k**-point [Fig. 2(b) and (c)]. The gauge must be recovered to that of the original and specific set of wavefunctions to correctly proceed with the Wannier interpolation of the *e*-ph matrix elements. In other words, the symmetry-unfolded *e*-ph matrix elements (complex numbers) must be exactly the same (up to numerical convergence accuracy) as those directly computed without using any symmetries, in both their magnitudes and phases.

To achieve this goal, we developed and implemented a symmetry unfolding technique for *e*-ph matrix elements with gauge recovering. This technique is implemented as an executable sympert.x in the BerkeleyGW software package and is suitable for both DFPT and *GW*PT *e*-ph matrix elements. It is designed to request the direct computation of *e*-ph matrix elements only within the symmetry-reduced **q**-grid (along with full **k**-grids for each **q**-point), and then the *e*-ph matrix elements on the full **q**-grid can be obtained by symmetry unfolding. The gauges of the unfolded matrix elements are recovered to be consistent with the original and specific set of wavefunctions $\{\psi_{n\mathbf{k}}\}$.

We first obtain a set of wavefunctions and denote their lattice-periodic parts of the Bloch wavefunctions as $u_{n\mathbf{k}}(\mathbf{r})$ where $\mathbf{k} \in$ full BZ. We define a symmetry operator $\{S|\mathbf{v}\}$ (which represents a symmetry of the crystal being studied) acting on vectors in the Cartesian coordinates,

$$\{S|\mathbf{v}\}\mathbf{r} = S\mathbf{r} + \mathbf{v}, \tag{13}$$

where $S$ is a 3×3 rotation matrix and **v** is a fractional lattice translation vector. For each symmetry operation, we apply the symmetry operator to all the wavefunctions and obtain a new set of rotated wavefunctions $\tilde{u}_{nS\mathbf{k}}(\mathbf{r})$,

$$\tilde{u}_{nS\mathbf{k}}(\mathbf{r}) = e^{-iS\mathbf{k}\cdot\mathbf{v}} u_{n\mathbf{k}}(S^{-1}\mathbf{r} - S^{-1}\mathbf{v}). \tag{14}$$

At a given **k**, we now have two different $u(\mathbf{r})$'s – one from the direct full **k**-set calculation, and the other one rotated from a symmetry-related wavefunction at a different **k**-point. The gauge (phase) difference between the rotated wavefunction and the original directly computed wavefunction at **k** is $\langle \tilde{u}_{n\mathbf{k}}|u_{n\mathbf{k}}\rangle$ for non-degenerate states. We generalize this gauge difference by introducing an overlap matrix $D$ (for each symmetry operation), spanning over all bands of interest, such that all degenerate states can naturally be taken care of,

$$D_{mn}^{\mathbf{k}} = \langle \tilde{u}_{m\mathbf{k}}|u_{n\mathbf{k}}\rangle. \tag{15}$$

Now we look at the symmetry relation for atom positions. Each atom within the Born-von-Karman supercell can be located by its position vector,

$$\mathbf{x}(l, \kappa) = \mathbf{R}_l + \boldsymbol{\tau}_\kappa, \tag{16}$$

where $\mathbf{R}_l$ is a lattice vector for the *l*-th unit cell and $\boldsymbol{\tau}_\kappa$ is the coordinates within a unit cell of the *κ*-th atom. A symmetry-equivalent site $\mathbf{x}(L, K) = \mathbf{R}_L + \boldsymbol{\tau}_K$ under $\{S|\mathbf{v}\}$ (up to a lattice vector translation) can be related as [41],

$$\mathbf{x}(L, K) = \{S|\mathbf{v} + \mathbf{R}_m\}\mathbf{x}(l, \kappa) = S\mathbf{x}(l, \kappa) + \mathbf{v} + \mathbf{R}_m, \tag{17}$$

where $K$ labels a *κ*-equivalent atom and $\mathbf{R}_m$ is a lattice vector needed in general to fulfill the symmetry relation.

To perform the *e*-ph matrix elements symmetry unfolding, we first work in the Cartesian-coordinate basis with $\alpha, \beta = x, y, z$. Given that we have $g_{mn\kappa\alpha}(\mathbf{k}, \mathbf{q})$ on the irreducible **q**-wedge (with full **k**-grid), we can generate the *e*-ph matrix element for $S\mathbf{q}$ and the displacement perturbation $K\alpha$ with gauge recovering as [32],

$$g_{mnK\alpha}(\mathbf{k}, S\mathbf{q}) = e^{i\mathbf{q}\cdot\boldsymbol{\tau}_\kappa - iS\mathbf{q}\cdot\boldsymbol{\tau}_K + iS\mathbf{q}\cdot\mathbf{v}} \sum_{m'n'} \left(D_{m'm}^{\mathbf{k}+S\mathbf{q}}\right)^* D_{n'n}^{\mathbf{k}} \sum_\beta S_{\alpha\beta} g_{m'n'\kappa\beta}(S^{-1}\mathbf{k}, \mathbf{q}). \tag{18}$$

Lastly, we rewrite the above gauge-recovering symmetry-unfolding relation in crystal-coordinate basis as used by ABINIT and BerkeleyGW. We introduce the lattice vector matrix $A$ as,

$$A = (\boldsymbol{a}_1, \boldsymbol{a}_2, \boldsymbol{a}_3)^T, \tag{19}$$

and then the rotation matrix in the crystal-coordinate basis can be found as,

$$S^{crys} = (A^T)^{-1} S A^T. \tag{20}$$

Eq. (18) can then be rewritten in the crystal-coordinate basis as [32],

$$g_{mnKa}(\mathbf{k}, S\mathbf{q}) = e^{i\mathbf{q}\cdot\boldsymbol{\tau}_\kappa - iS\mathbf{q}\cdot\boldsymbol{\tau}_K + iS\mathbf{q}\cdot\mathbf{v}} \sum_{m'n'} \left(D_{m'm}^{\mathbf{k}+S\mathbf{q}}\right)^* D_{n'n}^{\mathbf{k}} \sum_b g_{m'n'\kappa b}(S^{-1}\mathbf{k}, \mathbf{q})[(S^{crys})^{-1}]_{ba}, \tag{21}$$

where $a, b = \boldsymbol{a}_1, \boldsymbol{a}_2, \boldsymbol{a}_3$ label the primitive unit cell lattice vectors. The expression of Eq. (21) and its implementation in sympert.x within BerkeleyGW allow us to perform direct *GW*PT (and DFPT) calculations only on the symmetry-reduced **q**-grid. Then the unfolded full **q**-grid *e*-ph matrix elements with gauge recovering can be interpolated by the Wannier functions whose gauge has been fixed to the original and specific set of wavefunctions $\{\psi_{n\mathbf{k}}\}$ on the full **k**-grid. Note that the above Eqs. (13) – (21) are defined for one symmetry operation $\{S|\mathbf{v}\}$. Typically, to unfold the irreducible **q**-BZ to full **q**-BZ, multiple symmetries are needed following the same procedure. This symmetry unfolding technique with gauge recovering dramatically reduces the computational expense, especially for *GW*PT.

**V. Example: Boron-doped diamond**

We demonstrate the workflow using an example of boron-doped diamond, B$_x$C$_{1-x}$ [3,4], a superconductor with experimental $T_c \sim 4$ K [42]. Previous DFT and DFPT calculations have shown that B-dopant derived phonon modes enhance the *e*-ph coupling [3]. Here, as a demonstration of the *GW*PT workflow, we neglect the disorder effect and adopt a virtual crystal approximation (VCA) [43] for B$_{0.0185}$C$_{0.9815}$ ($x = 0.0185$), working with a two-atom primitive unit cell. The calculations involve different sets of **k**-, **q**-, and **p**-meshes that are tabulated in Table I. Only the phonon **q**-mesh is symmetry reduced. Before interpolation, the **k**-, **q**-, and **p**-meshes are required to be commensurate such that all $(\mathbf{k} + \mathbf{q})$, $(\mathbf{k} - \mathbf{p})$, and $(\mathbf{k} + \mathbf{q} - \mathbf{p})$ points should exist in the **k**-mesh. Norm-conserving pseudopotentials are taken from the PseudoDojo library [44], and a wavefunction energy cutoff of 60 Ry is used in DFT and DFPT calculations with ABINIT. The screened Coulomb interaction is built with an energy cutoff of 15 Ry and 200 bands, and the self-energy operator and its first-order changes are constructed by summing over 200 bands using BerkeleyGW. The EPW code interpolates the band energies, phonon frequencies, and *e*-ph matrix elements to the fine **k**- and **q**-meshes.

Like DFPT calculations, *GW*PT calculations are carried out in the primitive unit cell, and typically the computational cost of a single-mode-$\mathbf{q}\kappa a$ *GW*PT calculation is similar to that of a single *GW* calculation. Therefore, full *GW*PT calculations of one material system can be a few orders of magnitude computationally heavier than *GW* due to the large number of phonon modes, i.e., $N_{\text{mode}} = 3 \times N_{\text{atom}} \times N_{\mathbf{q}}$. The advantage of linear-response approaches (used in DFPT and *GW*PT) is that all phonon modes are independent, and therefore the computational cost scales virtually linearly with $N_{\text{mode}}$ being considered. Fig. 3 shows the computational costs of several main steps in this B-doped diamond example. Here, the DFPT and *GW* calculations have similar computational costs, which are one to two orders of magnitude heavier than the DFT calculations. Comparing *GW* and *GW*PT, *GW*PT is around two orders of magnitude heavier than *GW*. This strong contrast shows the significant computational resources required for *GW*PT calculations.

With the workflow, we can compute various *e*-ph properties of B-doped diamond at the full *GW* level (including both the *GW* band energies and *GW*PT *e*-ph matrix elements). Fig. 4 shows the Wannier-interpolated DFT and *GW* band structures. States near the Fermi energy $E_F$ remain similar, despite an increased band width due to *GW* self-energy effects. Fig. 5 shows the calculated Eliashberg function $\alpha^2 F(\omega)$ at both the DFT and *GW* levels. The *GW* self-energy effects enhance the overall *e*-ph coupling strength, mostly from the renormalization in the *e*-ph matrix elements. We arrive at a DFT-level *e*-ph coupling constant of $\lambda^{\text{DFT}} = 0.228$, in good agreement with previous DFPT-VCA calculations with a value of 0.237 [3]. At the *GW* level, we arrive at a $\lambda^{GW} = 0.302$, showing a 32% enhancement compared with $\lambda^{\text{DFT}}$. Table II presents the superconducting $T_c$ analysis using the McMillan-Allen-Dynes formula [45,46]. Within a physical range of 0.05 – 0.1 for the effective Coulomb parameter $\mu^*$, the DFT superconducting $T_c$ is 0.65 – 0.015 K, and the *GW* superconducting $T_c$ is 4.2 – 0.69 K. The *GW* results show enhanced $T_c$ values due to correlation effects, in better agreement with experiments. Note that the phonon modes from disordered boron dopants can possibly further enhance the *e*-ph coupling strength [3], and its cooperative interaction with the *GW* renormalization effects is beyond the scope of this work.

## VI. Summary

We presented a practical workflow for *GW*PT calculations of *e*-ph couplings at the many-electron level. This workflow combines three *ab initio* software packages – namely, BerkeleyGW, ABINIT, and EPW. To reduce the computational cost, especially for *GW*PT, we have developed a gauge-recovering symmetry unfolding technique, fulfilling the gauge consistency requirement for Wannier interpolation. An example of B-doped diamond shows a 32% enhancement in the *e*-ph coupling constant near $E_F$ due to many-electron correlation effects. This workflow enables rich *e*-ph research opportunities based on *GW*PT.

**Acknowledgements**

This work was mainly supported by the National Science Foundation under Grant No. OAC-2103991. Advanced code for *GW* and *GW*PT calculations were provided by the Center for Computational Study of Excited-State Phenomena in Energy Materials (C2SEPEM) at LBNL, which is funded by the U.S. Department of Energy, Office of Science, Basic Energy Sciences, Materials Sciences and Engineering Division under Contract No. DEAC02-05CH11231, as part of the Computational Materials Sciences Program. Computational resources were provided by Frontera at the Texas Advanced Computing Center (TACC), which is supported by National Science Foundation under Grant No. OAC1818253.

**Table I.** Different wavevector meshes used in the B-doped diamond example. Only the coarse phonon **q**-mesh is symmetry reduced.

| Software | Procedure | Key quantity | **k**-mesh | **q**-mesh | **p**-mesh |
|---|---|---|---|---|---|
| ABINIT | DFT | $\psi_{n\mathbf{k}}$ | 8×8×8 | – | – |
| ABINIT | DFPT | $\Delta_{\mathbf{q}\kappa a}\psi_{n\mathbf{k}}$ $g_{mn\kappa a}^{\text{DFT}}(\mathbf{k},\mathbf{q})$ | 8×8×8 | 4×4×4 (8 irreducible **q**-points) | – |
| Wannier90 | Wannierization | $U_{nw\mathbf{k}}$ | 8×8×8 | – | – |
| BerkeleyGW | GW | $\epsilon_{\mathbf{GG}'}^{-1}(\mathbf{p})$ | 8×8×8 | – | 8×8×8 |
| BerkeleyGW | GW | $\langle\psi_{n\mathbf{k}}|\Sigma|\psi_{n\mathbf{k}}\rangle$ | 8×8×8 | – | 8×8×8 |
| BerkeleyGW | GWPT | $g_{mn\kappa a}^{GW}(\mathbf{k},\mathbf{q})$ | 8×8×8 | 4×4×4 (8 irreducible **q**-points) | 8×8×8 |
| EPW | Interpolation | $g_{mn\nu}(\mathbf{k}_{\text{fi}},\mathbf{q}_{\text{fi}})$ | 40×40×40 | 20×20×20 | – |

**Table II.** Calculated *e*-ph coupling constant $\lambda$, logarithmic-averaged phonon frequency $\omega_{\log}$, and McMillan-Allen-Dynes superconducting transition temperature $T_c$ with different values of effective Coulomb potential parameter $\mu^*$.

|  | $\lambda$ | $\omega_{\log}$ (K) | $T_c$ (K) | | |
|---|---|---|---|---|---|
|  |  |  | ($\mu^* = 0.05$) | ($\mu^* = 0.08$) | ($\mu^* = 0.1$) |
| DFT + DFPT | 0.228 | 1360.6 | 0.65 | 0.099 | 0.015 |
| *GW* + *GW*PT | 0.302 | 1348.7 | 4.2 | 1.6 | 0.69 |

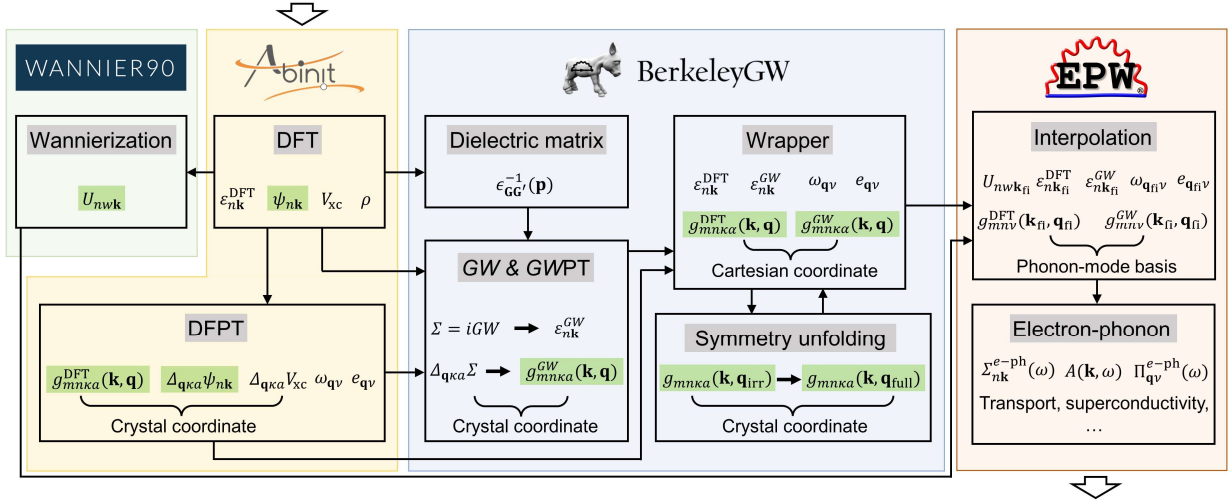

**Fig. 1.** Practical workflow for *GW*PT calculations, combining BerkeleyGW, ABINIT, and EPW software packages. The workflow starts with DFT and DFPT calculations using ABINIT, along with Wannierization using Wannier90. The *GW* and *GW*PT calculations are performed using BerkeleyGW, which includes a wrapper step which performs elph_interface and gauge-recovering symmetry unfolding (for the **q**-grid) of *e*-ph matrix elements using sympert.x. The DFT-level and *GW*-level electron states and *e*-ph matrix elements are then taken by EPW for interpolation and computation of *e*-ph properties on fine **k**- and **q**-grids. Quantities highlighted by the green boxes are gauge-consistent with each other and are fixed to the specific gauge of the set of wavefunctions $\{\psi_{n\mathbf{k}}\}$ in the DFT step.

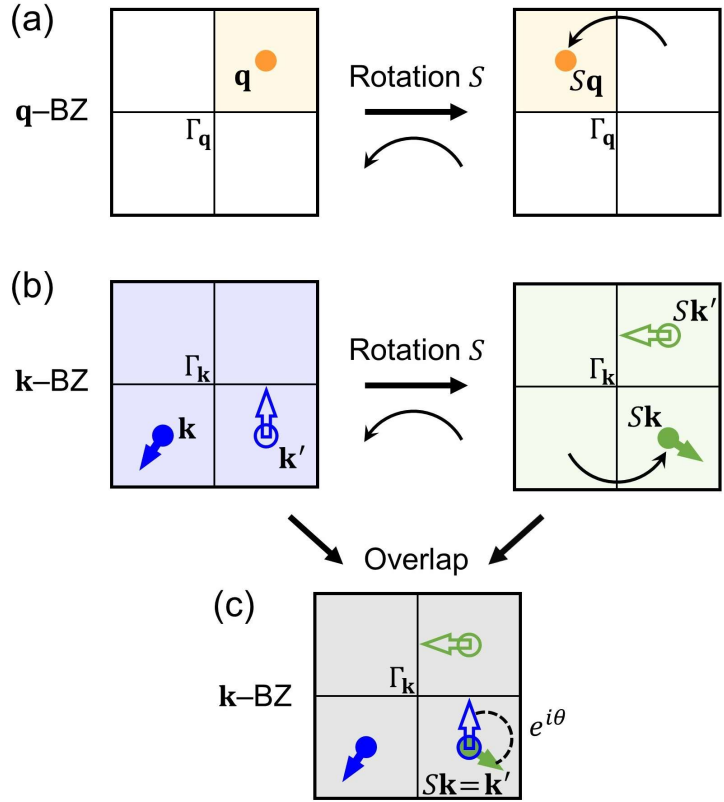

**Fig. 2.** Illustration of the gauge-recovering symmetry unfolding scheme. (a) Phonon **q**-mesh is symmetry-reduced, where the irreducible wedge is represented by the light orange shadow. With a rotation operation $S$, the **q**-BZ can be unfolded from the irreducible part to another part. (b) Electron **k**-BZ is always populated with directly computed electron wavefunctions in the full BZ in this scheme. Each wavefunction acquires a random gauge (phase) from solving an eigen-equation. The gauge is illustrated as an arrow at every **k**-point. Under the same symmetry operation $S$, a new set of rotated wavefunctions is generated. (c) Comparing the wavefunction at the same $S\mathbf{k} = \mathbf{k}'$ point from the unrotated and rotated sets, a gauge difference $e^{i\theta}$ can be found. Since all $e$-ph matrix elements depend on the gauge of the wavefunction of both the initial and final states, by computing the gauge differences, the gauge of the $e$-ph matrix elements can be recovered to that derived from the original specific set of wavefunctions $\{\psi_{n\mathbf{k}}\}$.

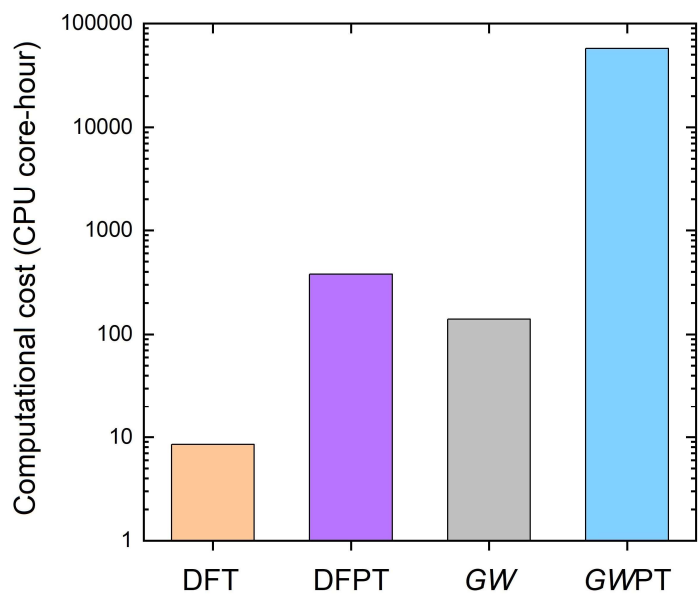

**Fig. 3.** Computational costs of the DFT, DFPT, *GW*, and *GW*PT steps for the B-doped diamond example. Calculations were performed for a two-atom primitive unit cell, using Intel 8280 "Cascade Lake" CPUs (56 cores per node) on Frontera at the Texas Advanced Computing Center.

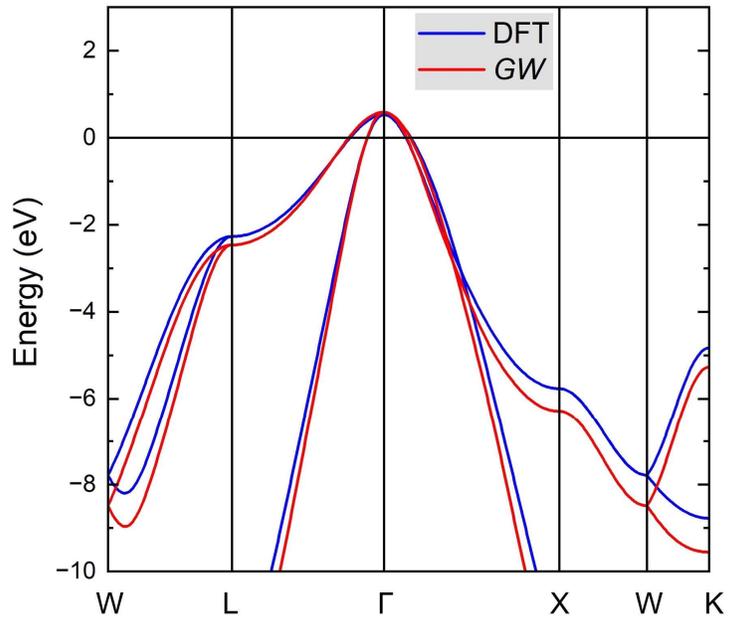

**Fig. 4.** DFT and *GW* band structures of B-doped diamond with 1.85% B-dopant concentration. The Fermi energy $E_F$ is set to 0 eV.

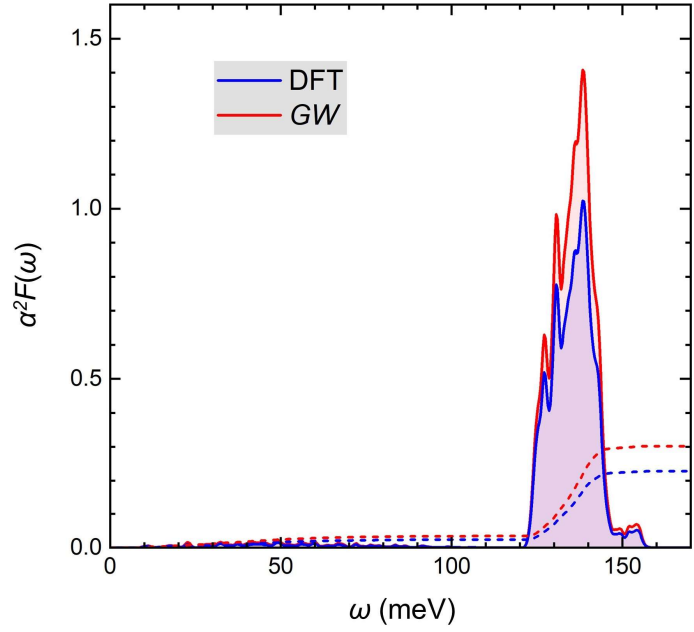

**Fig. 5.** Eliashberg function $\alpha^2 F(\omega)$ computed at DFT (blue) and *GW* (red) levels. The dashed lines represent the running integral of $\lambda^<(\omega) = 2 \int_0^\omega \frac{\alpha^2 F(\omega')}{\omega'} d\omega'$. The fully-integrated *e*-ph coupling constants are $\lambda^{\text{DFT}} = 0.228$ and $\lambda^{GW} = 0.302$.